\documentclass[10pt, conference]{IEEEtran}
\usepackage{booktabs}
\usepackage{array}
\usepackage{lipsum}
\usepackage{graphicx}
\usepackage{amsmath}
\usepackage{amssymb}
\usepackage{xcolor}
\usepackage{xfrac}
\usepackage{braket}

\usepackage{subcaption}
\usepackage{comment}
\usepackage{float}
\usepackage{mathtools}
\usepackage{tikz}
\usetikzlibrary{quantikz2}
\usepackage[]{mdframed}
\usepackage{multirow}
\usepackage{algorithm}
\usepackage{algpseudocode}
\usepackage{soul}
\usepackage{hyperref}

\usepackage{cite}
\usepackage{amsmath,amssymb,amsfonts}
\usepackage{graphicx}
\usepackage{textcomp}
\usepackage{xcolor}
\usepackage[inkscapelatex=false]{svg}

\usepackage[commandnameprefix=ifneeded]{changes}
\definechangesauthor[name={Mariya}, color=blue]{MB}

\newcommand{\vect}[1]{\ensuremath{\mathbf{#1}}}

\def\BibTeX{{\rm B\kern-.05em{\sc i\kern-.025em b}\kern-.08em
    T\kern-.1667em\lower.7ex\hbox{E}\kern-.125emX}}

\begin{document}
\title{Credit Default Prediction with Projected Quantum Feature Models and Ensembles}

\author{
\IEEEauthorblockN{Andras Ferenczi}
\IEEEauthorblockA{\textit{American Express}\\
andras.l.ferenczi1@aexp.com
}
\and
\IEEEauthorblockN{Dagen Wang}
\IEEEauthorblockA{\textit{American Express}\\
}
\and
\IEEEauthorblockN{Mariya Bessonova}
\IEEEauthorblockA{\textit{American Express}\\
}
\and
\IEEEauthorblockN{Sutapa Samanta}
\IEEEauthorblockA{\textit{American Express}\\
}
\and
\IEEEauthorblockN{Todd Hodges}
\IEEEauthorblockA{\textit{American Express}\\
}
\and
\IEEEauthorblockN{John Hancock}
\IEEEauthorblockA{\textit{American Express}\\
}
\and
\IEEEauthorblockN{Guillermo Mijares Vilariño}
\IEEEauthorblockA{\textit{IBM Quantum}, \\ \textit{IBM Research}\\
guillermo.mijares@ibm.com
}
\and
\IEEEauthorblockN{Amol Deshmukh}
\IEEEauthorblockA{\textit{IBM Quantum}, \\ \textit{IBM Research}\\
}
\and
\IEEEauthorblockN{Mariana LaDue}
\IEEEauthorblockA{\textit{IBM Quantum}, \\ \textit{IBM Research}\\
}
\and
\IEEEauthorblockN{Girish Pillai}
\IEEEauthorblockA{\textit{American Express}\\
}
\and
\IEEEauthorblockN{Hilary Packer}
\IEEEauthorblockA{\textit{American Express}\\
}
}
\maketitle

\begin{abstract}
Accurate prediction of future loan defaults is a critical capability for financial institutions that provide lines of credit.  
For institutions that issue and manage extensive loan volumes, even a slight improvement in default prediction precision can significantly enhance financial stability and regulatory adherence, resulting in better customer experience and satisfaction. 
Datasets associated with credit default prediction often exhibit temporal correlations and high dimensionality. These attributes can lead to accuracy degradation and performance issues when scaling classical predictive algorithms tailored for these datasets.
Given these limitations, quantum algorithms, leveraging their innate ability to handle high-dimensionality problems, emerge as a promising new avenue alongside classical approaches. 
To assess the viability and effectiveness of quantum methodologies, we investigate a hybrid quantum-classical algorithm, utilizing a publicly available ``Default Prediction Dataset'' released as part of a third-party data science competition.
Specifically, we employ hybrid quantum-classical machine learning models based on projected quantum feature maps and their ensemble integration with classical models to examine the problem of credit card default prediction.
Our results indicate that the ensemble models based on the projected quantum features were capable of slightly improving the purely classical results expressed via a ``Composite Default Risk'' (CDR) metric. 
Furthermore, we discuss the practical applicability of the studied quantum-classical machine learning techniques and address open questions concerning their implementation.
\end{abstract}

\begin{IEEEkeywords}
quantum machine learning, projected features, classification, credit default prediction
\end{IEEEkeywords}

\section{\label{sec:intro}Introduction}
Predicting consumer loan defaults is a critical capability for credit card issuers, given the growing prominence of credit cards as the de facto payment instrument in modern commerce. 
Banks currently utilize a range of models to identify credit default risk, though they don’t achieve perfect accuracy. 
This highlights the necessity for ongoing algorithm improvements and the exploration of novel technologies.

Traditional classical algorithms often face challenges with scalability and performance when handling the high-dimensional, temporally correlated data prevalent in the industry. 
Quantum computing emerges as a promising technology, offering the potential improvements in accuracy, efficiency and cost-effectiveness when combined with classical approaches. Reconceptualizing commercial algorithms in the quantum age is essential. 

In this paper, we examine a specific model architecture of hybrid quantum-classical machine learning model for the task of credit default prediction using an industrial scale dataset.
In general, hybrid quantum-classical machine learning algorithms exemplify a paradigm that utilizes quantum computational resources for feature embedding and classical optimization techniques for learning tasks. The model fundamentally depends on the encoding of classical datasets into parametric quantum feature maps, followed by the extraction of quantum expectation values that act as inputs to classical machine learning algorithms.

This document is organized as follows.
In Section~\ref{sec:survey} we survey the existing literature on credit default, and in general, binary classification tasks.
In Section~\ref{sec:background}, we outline the data and the evaluation framework.   
The modeling background is specified in Section~\ref{sec:QML}. 
We discuss the results in section~\ref{sec:results}.
In Section~\ref{sec:conclusion}, we examine the theoretical conclusions including runtime complexity, robustness, and potential expansions of the method.


\section{\label{sec:survey}Previous research survey}
Credit card default prediction is a binary classification task which is ubiquitous in machine learning and has been studied extensively in the literature. 
Various classical ensemble models that combine multiple models have been proposed to improve the classification ability. 
Please see~\cite{rokach2010} for a comprehensive review. 
An ensemble of XGBoost, Random Forest and the TabNet deep learning model showed a promising overall improvement in fraud detection~\cite{Xing2024EnhancedCS}.

Quantum classifiers hold a promise to improve performance compared to their classical counterparts. 
There exist several theoretical proposals combined with experiments on classical simulators and quantum hardware for realizing faithful quantum classifiers. 
Classical data encoding methods, a non-trivial challenge for QML, were discussed in~\cite{schuld2019quantum, P_rez_Salinas_2020, Schuld:2018ahn}. 
Quantum kernel based classifiers are studied in~\cite{Blank:2020zfq,Park_2020,Alvarez-Estevez:2024avx}. 
Small-scale experiments were performed on classical simulators with a noise model. 
Quantum-inspired hybrid neural network based classifiers were discussed in~\cite{Farhi:2018nhu, Tacchino_2019, liao2019quantumadvantagetrainingbinary}. 
The implementations were restricted to classical simulators or very small scale quantum hardware. 
In ~\cite{Rodriguez_Grasa_2025}, a combination of the quantum neural network and the projected quantum kernel was proposed.
Algorithm stability in the presence of noise was demonstrated using classical simulations with a noise model. 
A hybrid quantum-classical neural network model suitable for binary classification of noisy data was shown to perform better than existing quantum classifiers and on par with classical classifier in~\cite{Schetakis:2022aaw}. 
Various other quantum-inspired algorithms for classification have also been proposed, e.g., a  variational quantum classifier ~\cite{Cappelletti_2020}, a quantum probability theory-based algorithm ~\cite{Tiwari2019}, and a distance-based classifier using quantum interference circuit~\cite{Schuld_2017}. 
A hybrid classical-quantum deep learning model for credit risk assessment was developed in~\cite{minati2025quantumpoweredcreditrisk}. 
Another credit scoring approach called systematic quantum scoring (SQS)~\cite{Mancilla:2024tyv}, and preliminary experiments on a classical simulator showed improved performance over XGBoost. 
Quantum Support Vector Machine (QSVM) was applied for credit card fraud detection in~\cite{Grossi:2022gdj}. 
The experiments were performed primarily on a classical simulator. 
The experiment conducted on a quantum device was limited to a small subset of the data. 
Quantum multiple kernel learning was utilized for credit card data classifications in~\cite{vedaie2020quantummultiplekernellearning, miyabe2023quantum}. 
While the experiments conducted in~\cite{vedaie2020quantummultiplekernellearning} were performed using classical simulations, those in~\cite{miyabe2023quantum} utilized quantum hardware, involving up to 20 qubits.
Comparable analyses were conducted on the "Default Prediction Dataset", as detailed in~\cite{khurana2025integration}.


\section{\label{sec:background}Background}

\subsection{\label{sec:background_data}Data and evaluation criteria}

We utilize a publicly available ``Default Prediction Dataset''~\cite{amex-default-prediction} released as part of  a third-party data science competition. 
The dataset consists of anonymized credit card customer profiles, with approximately 5 million monthly records for around 459,000 customers. Each record includes delinquency, spending, payment, balance, and risk indicators, along with time-series behavioral information.
The objective is to predict if a customer will default on their balance in the next 18 months.
Performance is evaluated on an out of time dataset containing $924,621$ customers. 
The dataset is characterized by substantial proportions of missing values, and pronounced class imbalance, necessitating advanced preprocessing strategies and robust modeling techniques. 
Its considerable size poses additional computational challenges, rendering it a benchmark for scalable machine learning approaches in financial risk assessment.

%
Our methodology begins by selecting a class-balanced or stratified subsample from the credit default dataset, removing non-modeling variables, discarding zero-variance features and splitting the data into training and test sets of the same size. Consequently, throughout this work, a dataset with $n_s$ samples refers to both the training and test set containing $n_s$ samples each, unless stated otherwise. 

Performance is evaluated using a ``Composite Default Risk''  (CDR) metric, originally defined for the same public competition. This metric is calculated as the mean of the normalized Gini coefficient and the default capture rate at $4\%$, with negative labels weighted by $20$ to address class imbalance. 


\section{Details of the hybrid quantum-classical algorithm}\label{sec:QML}
This section introduces quantum machine learning (QML) and defines the algorithm specific components pertinent to this work. 
We subsequently outline the comprehensive modeling workflow employed in this research.

\subsection{Quantum Machine Learning}
Quantum Machine Learning (QML) is an emerging interdisciplinary field that integrates quantum computing with machine learning, focusing on the application of quantum mechanical principles~\cite{schuld_petruccione_2021}. 
The overarching objective of QML is to attain a practical quantum advantage in data-driven decision analytics, similar to that of the entire field of quantum computing~\cite{lanes2025framework, huang2025vast}.
In general, a quantum advantage denotes the execution of an information processing task on quantum hardware that satisfies two essential criteria:
i) the correctness of the output can be rigorously validated, and
ii) it is performed with a quantum separation that demonstrably offers superior efficiency, cost-effectiveness, or accuracy compared to what is attainable through classical computation alone~\cite{ibm_quantum_advantage}.
Quantum systems possess characteristics such as superposition, entanglement, and interference, which can theoretically be utilized to represent and process information in ways that go beyond classical limitations.

Recent advances underscore two complementary trajectories. 
Firstly, quantum algorithms for linear algebra, optimization, and sampling indicate the potential for polynomial or even exponential acceleration in specific subroutines essential to machine learning~\cite{PhysRevLett.113.130503, liu2021rigorous, glick2024covariant, Jäger2023, leone2022practical}. 
Secondly, QML as a learning paradigm is recognized for encompassing a broader array of efficiently learnable concept classes compared to classical machine learning framework.  
Quantum learners have exhibited the capacity to address learning tasks that are challenging to classical learners in specific highly structured problems. 
Nonetheless, attaining a true quantum advantage characterized by a thoroughly confirmed and verifiable enhancement over classical methods continues to be an unresolved and formidable goal.
Critical challenges in QML encompass the effective encoding of classical information into quantum states, the reduction of hardware noise, and the meticulous evaluation of proposed algorithms against swiftly evolving classical techniques, among others. 
Recent studies~\cite{PhysRevLett.113.130503, liu2021rigorous, glick2024covariant, Jäger2023, leone2022practical} suggest that quantum machine learning may ultimately provide computing advantages and innovative modeling frameworks. 
Consequently, QML signifies a promising although still emerging field in the overarching endeavor to include quantum computation into practical information processing.

Furthermore, QML algorithms can be broadly divided into those that require fault-tolerant quantum hardware and those designed for near-term error-mitigated devices~\cite{Preskill_2018, kim2023scalable}. 
Our focus is on the latter, particularly hybrid quantum-classical methods framed within variational quantum algorithmic paradigm~\cite{Peruzzo2014}. 
These rely on parameterized quantum circuits (PQCs) in which quantum devices prepare states encoding data and, in some cases, model parameters, whose observables are extracted through measurements and processed classically, either directly or through iterative feedback. 
While the variational framework is well suited to noise resilience, it primarily produces heuristic algorithms with unclear asymptotic guarantees. 
Nevertheless, reflecting the historical trajectory of classical machine learning, current research emphasizes empirical demonstrations and benchmarking of heuristic approaches as hardware scales~\cite{bowles2024better}. 
Ultimately, tangible quantum advantages in QML are expected to depend on both the structure of the data and the alignment between the inductive biases of quantum models and the target learning tasks~\cite{kubler2021inductive, gili2023inductive}.

\subsection{Algorithm specific components}

In this section, we outline the fundamental components of the QML architectures employed in this work. 
As discussed earlier, QML on pre-fault-tolerant devices is predominantly realized through variational quantum computing models~\cite{Peruzzo2014}.
Most of these near-term quantum machine learning algorithms rely on a \textit{feature map}~\cite{schuld2019quantum}, introduced in Section~\ref{sec:fm}, which embeds classical data into quantum states. 

\subsubsection{Feature Maps}\label{sec:fm}
In classical machine learning, feature maps\cite{schuld2019quantum, schuld_petruccione_2021} embed data $\vect{x} \in \mathcal{X}$ into a transformed space $\mathcal{F}$, often enabling linear models to succeed on data that is not linearly separable in its original representation~\cite{Cristianini_Shawe-Taylor_2000}.
In quantum machine learning, analogously, quantum feature maps (QFMs) $\phi_Q: \mathcal{X} \rightarrow \mathcal{H}_Q$ encode classical inputs into the $2^n$-dimensional Hilbert space $\mathcal{H}_Q$ of an $n$-qubit device:
\begin{equation}
  \vect{x} \mapsto \phi_Q(\vect{x}) := |\vect{x}\rangle = U_{\phi_Q}(\vect{x})|0^n\rangle, 
\end{equation} 
whereby classical features $\vect{x}$ are mapped to quantum states $|\vect{x}\rangle$ via the action of a unitary quantum circuit $U_{\phi_Q}(\vect{x})$ on a convenient initial state such as the all-zero state of the $n$-qubit device $|0^n\rangle$.

In order to guarantee the applicability of these techniques to near-term devices, it is essential to modify the circuit according to the hardware architecture concerning qubit layout and gates. 
In addition, the use of error mitigation measures is recommended to reduce the impact of noise on measurement outcomes.

\subsubsection{\label{sec:Heisenberg}Heisenberg feature map}
Since quantum machine learning (QML) relies on quantum circuits, an important consideration is the classical non-simulability of the underlying ansatz, which necessitates genuine quantum hardware. To enforce this property, Hamiltonian simulation-based ansatzes~\cite{schuld_petruccione_2021} have been proposed. 
In this approach, functions of the form
$f(\vect{x})=\langle \vect{x}|e^{+iHt} O e^{-iHt}|\vect{x}\rangle$
with $H$ a many-body Hamiltonian, $O$ an observable, and $|\vect{x}\rangle$ the quantum feature map for input $\vect{x}$, are generally intractable to simulate classically. 
The simulability of the evolution operator $e^{iHt}$ depends intricately on the parameters $\beta_s$ in the decomposition $H=\sum_s \beta_s H_s$~\cite{harrow2017quantum, huang2021power}, which correspond to rotation angles in the circuit implementation. 
In this work, we employ the Heisenberg feature map, derived from the one-dimensional Heisenberg model.

The quantum state generated by the Heisenberg feature map for the embedding of classical data is
\begin{equation}\label{heisenberg_feature_map}
|\mathbf{x}_i \rangle = \left(\prod^{n-2}_{j=0} \exp\left(-i\frac{t}{T} H_j(x_{ij})\right) \right)^T \bigotimes^{n-1}_{j=0}|\psi_j\rangle,
\end{equation}
where
\begin{equation}
 H_j(x_{ij}) = x_{ij} (X_jX_{j+1} + Y_jY_{j+1} + Z_jZ_{j+1}),
\end{equation}
\noindent $x_{ij}$ are data values, $\{X,Y,Z\}$ are Pauli operators, $t$ is a time parameter, $T$ is the number of Trotter steps and $|\psi_j\rangle$ are fixed, Haar-random single-qubit states prepared at the beginning of the circuit. 
On quantum hardware, these Haar-random states are implemented by applying single-qubit unitaries sampled from the Haar measure and decomposed into native basis gates such as $RZ$, $\sqrt{X}$, and $X$ via transpilation.
Implementation in quantum circuits is done in a form that is factored into even- and odd-qubit indices.
\begin{equation}
|\mathbf{x}_i \rangle = \left(\prod_{\substack{j = 0 \\ j \text{ odd}}}^{n-2} U_j(x_{ij}) \prod_{\substack{j = 0 \\ j \text{ even}}}^{n-2} U_j(x_{ij}) \right)^R \bigotimes^{n-1}_{j=0}|\psi_j\rangle,
\end{equation}


\noindent where $U_j(x_{ij})=\exp(-i\alpha  x_{ij} (X_jX_{j+1} + Y_jY_{j+1} + Z_jZ_{j+1}))$, $R$ denotes the number of repetitions of circuit blocks, while $\alpha$ represents a temporal parameter that scales the data values, which can be understood as $t/T$ from the equation~\eqref{heisenberg_feature_map}. 
The final two variables are hyperparameters of the model.

\begin{figure}[h!]
    \centering
    \includegraphics[width=0.95\linewidth]{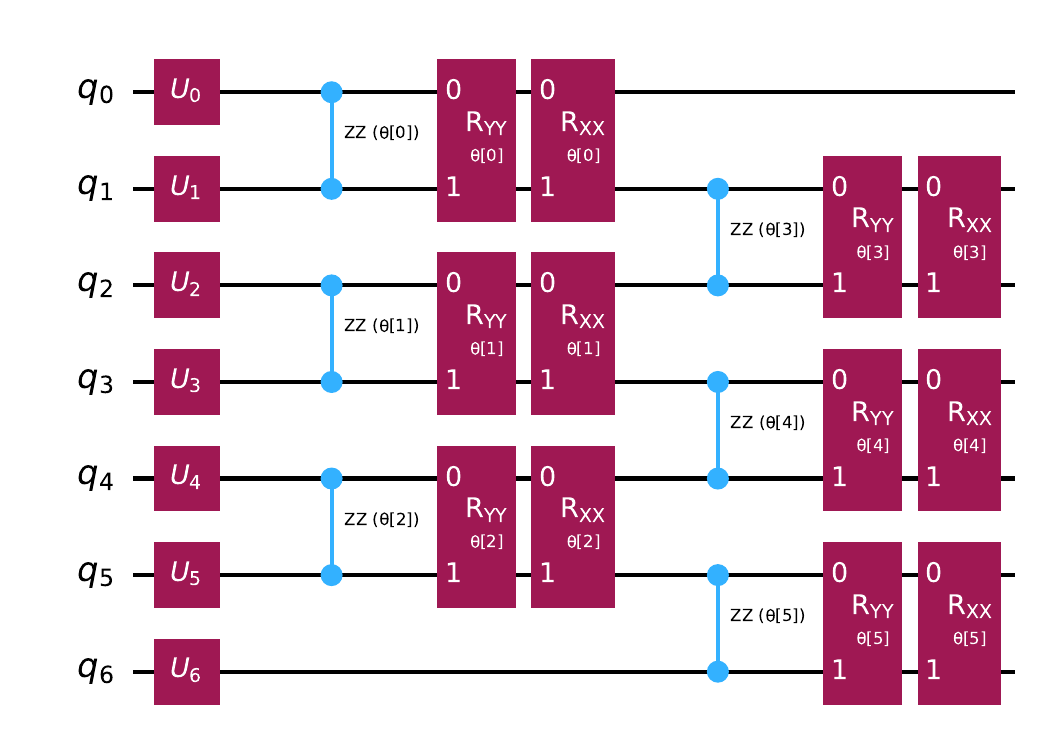}
    \caption{Example of a quantum Heisenberg feature map circuit with $7$ qubits encoding $6$ features and $1$ repetition. $U_i$'s are single-qubit Haar-random unitaries.}
    \label{fig:Heisenberg feature map}
\end{figure}

A quantum machine learning (QML) model implemented with a Heisenberg feature map tailored to the underlying structure of the task~\cite{huang2021power} generally exhibits computational complexity that precludes efficient classical simulation, thus necessitating quantum computational resources. 
Nevertheless, the realization of quantum-enhanced predictive performance is contingent upon the specific structural attributes and computational hardness of the learning problem~\cite{liu2021rigorous, glick2024covariant, leone2022practical}.

\subsubsection{\label{sec:PQF}Projected Quantum Features}

\begin{figure*}[t]
    \centering
    \includegraphics[width=\linewidth, scale = 1]{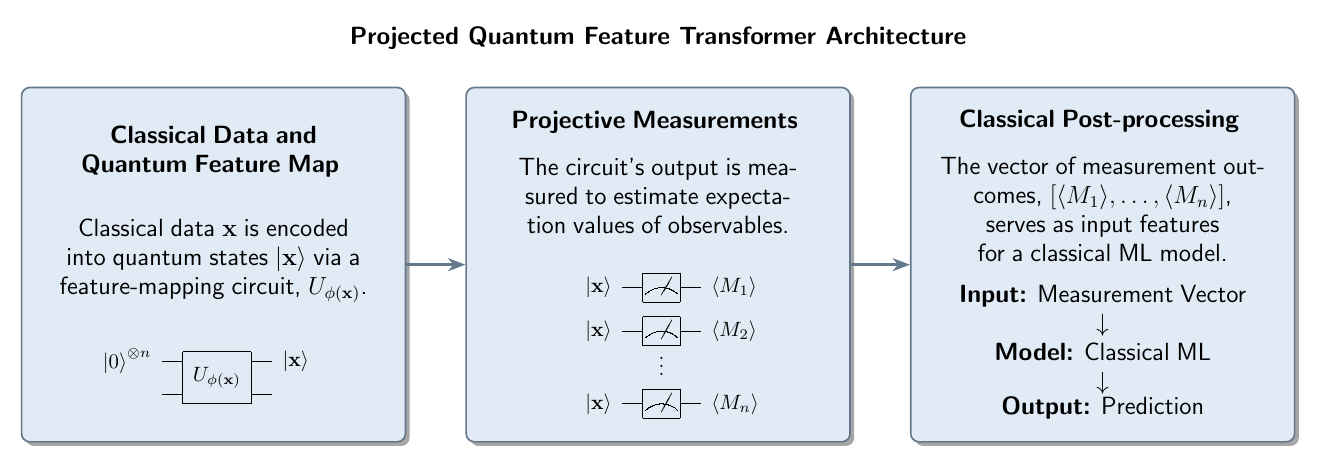}
    \caption{Diagram illustrating the projected quantum features (PQF) transformer.  
    Note this can, for example be used in a pipeline, preceded by normalization and followed by a machine learning model.}
    \label{fig:pqf}
\end{figure*}    

\textit{Projected Quantum Features (PQF)}~\cite{Huang2020classicalshadows, huang2021power} provide a framework for extracting classical feature vectors from quantum states via projective measurements. 
Data vectors $\vect{x}_i$ are first encoded through a quantum feature map circuit $U_{\phi_{Q}}(\vect{x}_i)$ applied to a fiducial state (e.g., $|0^n\rangle$), yielding $|\vect{x}_i\rangle = U_{\phi_{Q}}(\vect{x}_i)|0^n\rangle$. 
A chosen set of measurements $\mathcal{M}$ is then performed, producing outcomes $\vect{z}_i = \{m_1, \ldots, m_{|\mathcal{M}|}\} \in \mathcal{X}_Q$, which serve as classical features for downstream classical machine learning models. The architecture of the PQF transformer is illustrated in Fig.~\ref{fig:pqf}.
There is freedom to select the feature map $\phi_{Q}$, tailoring the PQFs to the requirements of the subsequent algorithm.

PQFs offer several advantages over fidelity-based quantum kernel methods:  
\begin{itemize}
    \item \textbf{Resource efficiency.} Fidelity kernels require evaluating pairwise overlaps $\langle \vect{x}_i | \vect{x}_j \rangle$, demanding quadratically many quantum device calls in dataset size. 
    PQFs instead generate classical vectors $\vect{z}_i$ per input, reducing quantum invocations to linear cost. 
    If desired, kernels can still be built post hoc using the PQF-derived features, shifting quadratic computation to classical post-processing.

    \item \textbf{Distinctive kernel structures.} Kernels derived from PQFs, known as \textit{Projected Quantum Kernels} (PQKs), can exhibit large geometric distance $g(K_C||K_Q)$~\cite{huang2021power} relative to classical kernels, a necessary condition for quantum advantage. 
    A notable example uses single-qubit one-body reduced density matrices (1-RDMs) $\rho_k(\vect{x}_i)$, obtained from local Pauli expectations $\{\langle X_k\rangle, \langle Y_k\rangle, \langle Z_k\rangle\}$ of the state $\rho(\vect{x}_i)$. 
    A simple yet powerful PQK is the Gaussian kernel on 1-RDM space:  
    \[
        \kappa^{PQ}(\vect{x}_i,\vect{x}_j) = \exp\!\left(-\gamma \sum_k \|\rho_k(\vect{x}_i) - \rho_k(\vect{x}_j)\|_F^2\right),
    \]  
    where $\gamma$ is a tunable parameter.

    \item \textbf{Noise robustness.} Empirical evidence suggests PQF based models retain predictive power under quantum hardware noise, making them more practical for near-term devices.  
\end{itemize}

\subsubsection{\label{sec:transpilation}Transpilation}
Transpilation is the process of adapting an ideal quantum circuit to the specifications of the real quantum hardware, including considerations for qubit connectivity and quantum gates usage. 
This process also entails optimizing these gates to reduce circuit depth and minimize the impact of noise as much as possible~\cite{ibm_transpiler_api, ibm_transpilation_guide}.

In this work, a preset pass manager from Qiskit~\cite{qiskitTranspilerPreset2025} with optimization level 3 was employed to adapt the quantum circuits to the quantum hardware. 
This pass manager searches for a perfect layout (i.e., one that satisfies all 2-qubit interactions) unless one is provided by the user. 
If the search is unsuccessful, and hardware calibration data is available, the circuit is mapped to the qubits and $CX$ gates with lowest error rates.
Subsequently, the circuit is transformed to comply with the device coupling map, unrolled to the target gate basis, and any incorrect $CX$ gate directions are corrected. 
The final optimization stage includes cancellation of commutative gates, resynthesis of 2-qubit unitary blocks, and removal of redundant reset operations.

\subsubsection{\label{sec:error_mitigation}Error mitigation}
Error mitigation techniques refer to methods that aim to reduce the effect of noise in quantum hardware without the need of error correction. 
Rather than preventing errors, these techniques model the noise characteristics of the quantum system and use post-processing strategies to extrapolate or infer ideal, noise-free outcomes from noisy measurements~\cite{steffen2022error, IBMQErrorMitigation2025}.

Experiments in this work employed the error mitigation model-free technique Twirled Readout Error eXtinction (TREX)~\cite{van_den_Berg_2022, IBMQErrorMitigation2025}. 
This technique consists of randomly applying X gates before measurements and inverting their effect through classical post-processing, effectively symmetrizing the readout noise. 
As a result the readout error matrix becomes diagonal and can be corrected classically. 
To characterize the noise, sampling is performed both on the target circuit and on an empty circuit preparing the $|0\rangle^{\otimes n}$ state.

\subsection{\label{sec:end_to_end}End-to-end architecture}

Our analysis leverages three distinct model categories: classical, quantum, and ensemble models, the latter combining the former two methodologies. 
The classical models are based on the gradient boosting-based classifier known as \texttt{XGBClassifier} from the Python version of the XGBoost library~\cite{chen2016xgboost, friedman2001greedy}.
The quantum models consist of pipelines that involve the following steps:
\begin{itemize}
    \item \texttt{MinMaxScaler} from \texttt{scikit-learn}~\cite{scikit-learn} that scales data features to a specified range, in our case $[0.3, 0.8]$. 
    This approach mitigates the issue of generating equivalent angles in quantum feature maps for distinct data values, which arises from the inherent periodicity of quantum gate parameters
    \item \texttt{ShuffleFeatures}, a data transformer that randomly permutes the feature order, thereby altering the mapping of features to rotation angles in the quantum feature map.
    \item \texttt{ProjectedQuantumFeatures}, a data transformer that projects a classical dataset into a quantum-enhanced feature space by applying a quantum feature map and computing expectation values of selected observables, thereby generating a new classical representation, as shown in Fig.~\ref{fig:pqf}. 
    This transformed dataset encodes structural properties of the quantum state, enabling the discovery of higher-order correlations and potentially improving model diversity and predictive performance.
    \item \texttt{XGBClassifier}, a classifier from the XGBoost library~\cite{chen2016xgboost}.
\end{itemize}

To obtain suitable hyperparameters for both the classical and the quantum models, we perform cross-validation with Bayesian optimization with different random seeds. 
Subsequently, the prediction probabilities of each pair of classical and quantum models were used to train ensemble models. 
These ensemble models primarily fall into two categories: mean-based models where the prediction probability of each class is calculated as the average of the classical and quantum models, and models involving a meta-classifier trained using the probabilities of the classical and quantum models.

\section{\label{sec:results}Simulator and hardware results}

\subsection{\label{sec:classical_benchmarks}Classical benchmarks}
\begin{table*}
\centering
    \begin{tabular}{|c|c|c|c|c|c|c|c|}
    \hline
    \textbf{Model} & \textbf{CV capture rate} & \textbf{CV Gini} & \textbf{CV CDR metric} &\textbf{test capture rate} & \textbf{test Gini} & \textbf{test CDR metric}\\	
    \hline
Gradient Boosting & 0.6988 $\pm$ 0.0085 & 0.9294 $\pm$ 0.0020 & 0.8141 $\pm$ 0.0050 & 0.6992 & 0.9295 & 0.8143 \\
Dummy Classifier & 0.0408 $\pm$ 0.0033 & 0.0100 $\pm$ 0.0115 & 0.0254 $\pm$ 0.0071 & 0.0396 & 0.0012 & 0.0204 \\
\hline
\end{tabular}
\caption{Performance comparison of classical machine learning models on cross-validation and test datasets using the CDR metric, Gini coefficient and capture rate at $4\%$.}
\label{tab:classical_benchmark}
\end{table*}
We create a baseline of performance metrics using classical models on the dataset as specified in the Sec.~\ref{sec:background_data}. 
The reduced dataset with a single month of data consists of $210,876$ records.
$50\%$ was randomly designated for training, while $50\%$ was allocated for testing. 
During cross-validation, the split is configured to 5.  
The Table~\ref{tab:classical_benchmark} presents a comparative analysis of classical machine learning models evaluated on both cross-validation (CV) and test datasets using  Gini coefficient, capture rate at $4\%$ and their average, which is the CDR metric.  
Hyperparameter tuning was performed using Bayesian optimization where the parameter search space is too large.
Gradient Boosting achieves strong performance as measured by the CDR metric. 
This model also maintains similar effectiveness on the test data, with test performances metrics (i.e., Gini coefficient, capture rate at $4\%$ and CDR metric) close to their CV metrics, indicating good generalization. 
We also report baseline results from Dummy Classifier, which records the low scores across all evaluation criteria, as expected. 
It is important to note that the benchmark results were obtained using the complete one-month dataset with classical models; therefore, subsequent comparisons with the methods presented later using restricted dataset in this paper should be interpreted with caution.

\subsection{\label{sec:preliminary}Preliminary analysis with simulators}
We commenced small-scale studies utilizing Qiskit Aer's statevector simulator and the tensor-network based simulator known as the matrix product state (MPS) simulator to assess the effectiveness of the PQF approach. 
In this work, `smaller scale’ denotes systems comprising $10$ to $\sim$ $30$ qubits, a regime that remains tractable for full statevector simulation.
The results of the statevector simulators were in turn utilized to test the accuracy of the simulations with the MPS simulator.
Please note, for these preliminary experiments, we utilized balanced subsamples, making metrics such as accuracy and AUC more meaningful.
On the other hand, the CDR metric assumes an approximate class distribution of $80\%$ negatives and $20\%$ positives, unless weighting adjustments are applied.
We selected the most important features from the dataset using scikit-learn implementation of Random Forest to determine the importance of the features.  
The pipeline employed in these trials has been previously delineated in \ref{sec:end_to_end}. 
Specifically, Heisenberg feature maps, as delineated in subsection \ref{sec:Heisenberg}, were utilized in all tests. Unless otherwise specified, the number of repetitions $R$ of the Heisenberg feature map is set to $1$.
For each value of the number of qubits, we perform a grid search over the $\alpha$ parameter for the Heisenberg feature map, and the learning rate parameter for the XGBoost model.  
For each combination of $\alpha$, the Pauli rotation factor for the Heisenberg feature map and the learning rate build a model to train and test on the sample of credit default data with five-fold cross-validation.  

\begin{table*}[ht!]
\resizebox{\textwidth}{!}{
\begin{tabular}{|c|c|c|c|c|c|c|c|c|c}
\hline
\textbf{Backend} & \textbf{No. of qubits} & \textbf{No. of features} & \textbf{No. of samples} & $\alpha$ & $\eta$ & \textbf{Train Acc.} & \textbf{Test Acc.} & \textbf{Classical Test Acc.}\\
\hline
Statevector simulator              & 20  & 100 & 1,000 & 0.1 & 0.3 & 100\%   & 86.40\% & 89.10\% \\
\hline
MPS simulator                      & 23  & 30  & 1,000 & 0.5 & 0.5 & 90.40\% & 87.20\% & 89.90\% \\
MPS Simulator                      & 51  & 100 & 1,000 & 0.75 & 0.1 & 100\%   & 88.30\% & 89.00\% \\
MPS Simulator                      & 81  & 160 & 1,000 & 1 & 0.3 & 100\%   & 89.10\% & 88.40\% \\
MPS Simulator                      & 81  & 160 & 1,000 & 1 & 0.1 & 100\%   & 88.70\% & 88.40\% \\
MPS Simulator                      & 101 & 170 & 1,000 & 0.1 & 0.3 & 100\%   & 88.80\% & 89.50\% \\
MPS Simulator                      & 101 & 174 & 1,000 & 0.25 & 0.5 & 100\%   & 90.30\% & 90.40\% \\
\hline
\textit{ibm\_brisbane}             & 81  & 160 & 1,000 & 1 & 0.3 & 100\%   & 87.70\% & 88.70\% \\
\textit{ibm\_torino}               & 101 & 174 & 1,000 & 1 & 0.0156 & 98.8\%  & 88.90\% & 89.80\% \\
\textit{ibm\_marrakesh}            & 101 & 174 & 500   & 2 & 0.3 & 100\%   & 88.80\% & 89.60\% \\
\textit{ibm\_marrakesh}                      & 101 & 174 & 500 & 2 & 0.1716  & 96.20\% & 88.00\% & 89.60\% \\
\hline
\end{tabular}
}
\caption{Results of initial experiments on balanced data. 
Each row records the backend used, the number of qubits used, number of features, and the number of samples considered. 
The train and test accuracies are then compared against the classical test accuracies.}
\label{tab:inital_experiments}
\end{table*}  

The first experiment employed exact statevector simulation and involved $20$ qubits, $100$ features and a number of repetitions $R=2$.
The attained test accuracy was $86.4\%$, whereas the traditional model achieved $89.10\%$. 
Given that the experiment yielded comparatively high quantum and classical accuracies, we set the ground for increasing the number of qubits and features while still employing classical simulation. 
To achieve this objective, we utilized the MPS simulator, as mentioned above. 
We began with a smaller-scale experiment with 23 qubits and 30 features and, after checking the quantum accuracies remained high, we progressively increased the number of qubits and features.

First, we fixed the number of features to $30$ and increased the number of qubits to get a better understanding of how the model scaled with the number of qubits.
In particular, we tested the following values: the range 10 through 18 consecutively, and 20, 30, 40 and 60. A balanced sample of 1,000 instances from the dataset was used. 
In Table~\ref{tab:num-qubits-test}, we report the performance that an XGBoost \cite{chen2016xgboost} classifier achieves when evaluated on the test data in terms of Accuracy (Acc.), Precision (Prec.), Recall (Rec.) and F1 score (F1). 
It also contains the values of the $\alpha$ parameter from the Heisenberg feature map and the learning rate ($\eta$) of the XGBoost.

\begin{table}[ht]
    \centering
    \begin{tabular}{|c|c|c|c|c|c|c|c|}
    \hline
    \textbf{Qubits} & $\alpha$ & $\eta$ & \textbf{Acc.} & \textbf{Prec.} & \textbf{Rec.} & \textbf{F1} & \textbf{AUC} \\
    \hline
    10 & 0.75 & 0.10 & 0.8800 & 0.89 & 0.87 & 0.88 & 0.9276 \\
    11 & 0.75 & 0.10 & 0.8570 & 0.86 & 0.86 & 0.86 & 0.9291 \\
    12 & 0.75 & 0.50 & 0.8510 & 0.85 & 0.85 & 0.85 & 0.9264 \\
    13 & 1.00 & 0.01 & 0.8580 & 0.87 & 0.84 & 0.86 & 0.9212 \\
    14 & 0.50 & 0.10 & 0.8760 & 0.88 & 0.87 & 0.88 & 0.9311 \\
    15 & 1.00 & 0.10 & 0.8580 & 0.85 & 0.85 & 0.86 & 0.9285 \\
    16 & 0.75 & 0.10 & 0.8760 & 0.89 & 0.85 & 0.87 & 0.9361 \\
    17 & 1.00 & 0.10 & 0.8800 & 0.90 & 0.86 & 0.88 & 0.9394 \\
    18 & 0.75 & 0.50 & 0.8830 & 0.88 & 0.89 & 0.88 & 0.9360 \\
    20 & 1.00 & 1.00 & 0.8790 & 0.88 & 0.87 & 0.88 & 0.9369 \\
    30 & 1.00 & 0.10 & 0.8930 & 0.90 & 0.88 & 0.89 & 0.9413 \\
    40 & 1.00 & 0.10 & 0.8820 & 0.89 & 0.88 & 0.88 & 0.9418 \\
    60 & 0.75 & 0.75 & 0.8690 & 0.86 & 0.88 & 0.87 & 0.9366 \\
    \hline
    \end{tabular}
    \caption{Results for quantum models built with a simulator evaluated on the test data}
    \label{tab:num-qubits-test}
\end{table}


The MPS simulator enabled simulations of up to $101$ qubits under the constraint of one-dimensional entanglement. 
Given that the Heisenberg feature map employs pairwise entanglement, MPS provides an appropriate framework for validating our experimental results.
Following this validation, we transitioned to IBM quantum hardware for subsequent experiments.

\subsection{\label{sec:hardware}Experiments with quantum hardware}
\begin{figure}[ht!]
    \centering
    \includegraphics[width=1.0\linewidth]{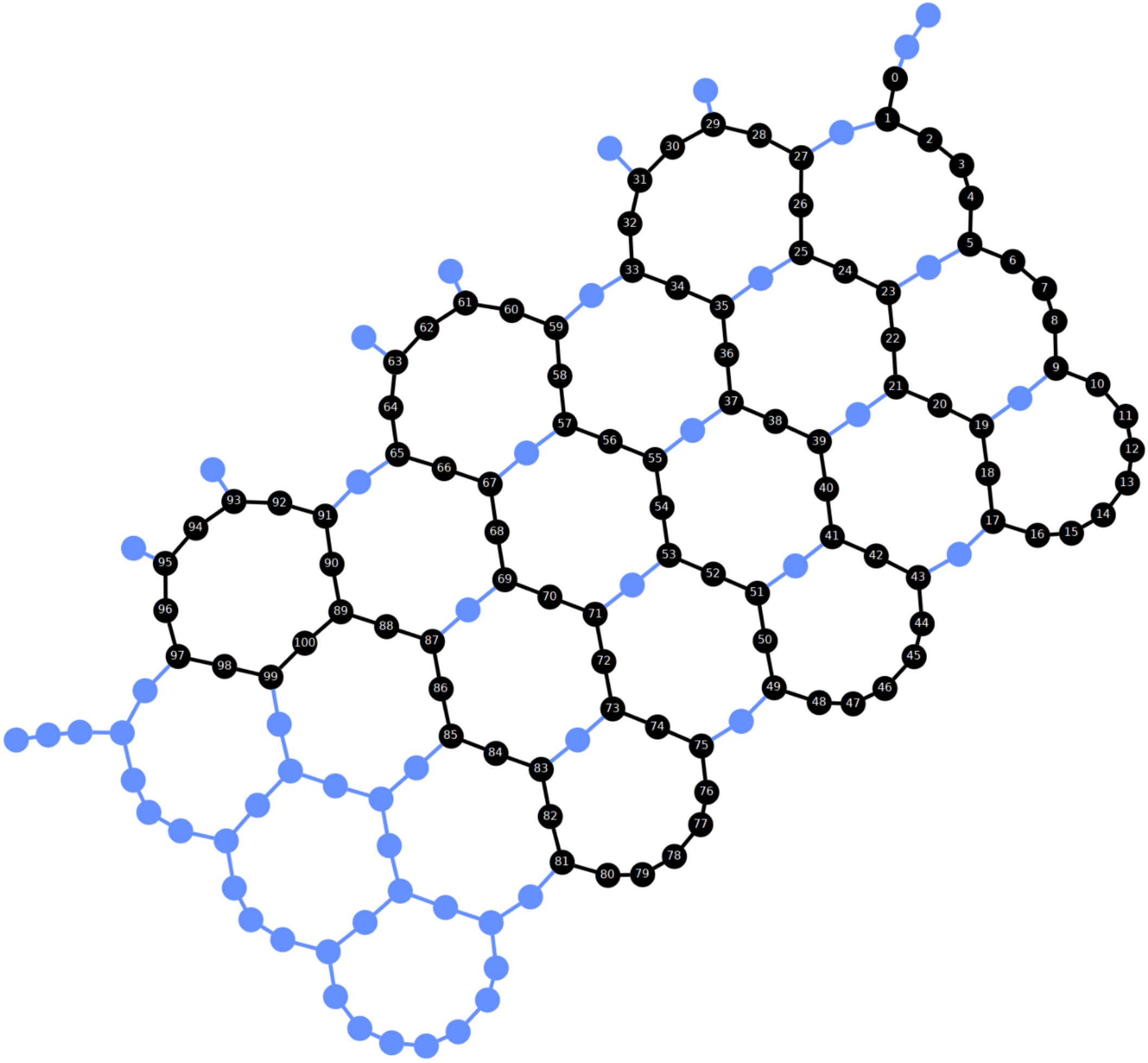}
    \caption{
    Topology of the quantum Heisenberg feature map circuit on \textit{ibm\_marrakesh} with 101 qubits encoding 174 features. 
    Filled circles denote qubits, and connecting lines indicate couplings. 
    Black circles and lines highlight the qubits and connections used in this experiment.
    }
    \label{fig:circuit layout}
\end{figure}

To maintain consistency with the simulator experiments from Section~\ref{sec:preliminary}, we kept using balanced data, enabling a fair comparison. The initial hardware experiment was conducted on \textit{ibm\_brisbane} utilizing $81$ qubits to encode $160$ features across $1,000$ samples. The transpiled Heisenberg feature map for this experiment had $870$ 2-qubit gates, a depth of $86$ and a 2-qubit depth of $20$. The latter refers to the circuit depth considering only the 2-qubit gates and it is particularly relevant for current noisy hardware, where the 2-qubit gates are far more error prone than single-qubit ones.
In this instance, the quantum model exhibited marginally lower test accuracy compared to the classical model, achieving $87.7\%$ as opposed to $88.7\%$ with the classical model. 
In the subsequent tests, we augmented the number of qubits to $101$ and the number of features to $174$ (including any feasible features in the classical dataset). 
The initial experiments under these conditions were conducted on \textit{ibm\_torino}, with a transpiled circuit that includes $870$ 2-qubit gates and has a depth of $86$ and a 2-qubit depth of $20$, yielding test accuracies of $88.9\%$ for the quantum model and $89.8\%$ for the classical model. 
The subsequent experiments were conducted on \textit{ibm\_marrakesh}, with transpiled circuit of depth $86$, 2-qubit depth of $20$, and 2-qubit gate count of $870$, utilizing $500$ samples instead of $1,000$. 
The quantum model achieved a test accuracy of $88.8\%$, while the conventional model attained $89.6\%$. 
The outcomes of our initial testing with balanced data are displayed in Table~\ref{tab:inital_experiments}.
Given that the results of all our hardware experiments, although lower than the classical results, were rather close, we augmented the sample size to $10,000$, initially in MPS and later in \textit{ibm\_marrakesh} after validating the MPS results. 
The analysis and results for the $10,000$-sample dataset are deferred until Section~\ref{sec:diversity_analysis}.
  
\subsection{\label{sec:diversity_analysis}Ensemble model and diversity analysis}


Ensemble modeling is the practice of combining multiple models, often with complementary strengths, to produce predictions that outperform any single base learner.
A typical ensemble consists of base learners (the individual models) and a rule for combining them (e.g., voting, averaging, or a meta-learner). 
Two common motivations for ensembling are: \textbf{variance reduction} — averaging over flexible but noisy models (e.g., decision trees in a random forest), and \textbf{bias reduction} — incrementally combining weak models that underfit individually (e.g., boosting).
Beyond these procedural ensembles, heterogeneous ensembles combine strong models from different families, each capturing distinct patterns in the data. 
For example, combining quantum and classical models leverages their complementary inductive biases and offers potential performance gains.

Diversity is a key factor in the construction of effective ensemble models~\cite{9915517}. 
The performance of an ensemble is determined not only by the accuracy of its base models but also by their diversity. 
When base learners make highly correlated errors, their combination yields limited benefit. 
In contrast, when models are individually accurate yet differ in their inductive biases or error patterns, the ensemble can offset weaknesses of individual members and achieve superior generalization.

Formally, ensemble generalization error can be expressed as an extension of the classic bias–variance decomposition \cite{wood2023unified}:
\begin{equation}\label{eq:bias_variance_diversity}
\text{expected loss} = \text{average bias} + \text{average variance} - \text{diversity}.
\end{equation}
Here, diversity quantifies the extent to which base models disagree (e.g., measured via correlation, covariance, or ambiguity \cite{zhouEnsemble2012}). 
A positive diversity term reduces ensemble error, making diversity a direct driver of ensemble performance.  

\begin{figure}[h!]
\centering
  \includegraphics[width=0.9\linewidth]{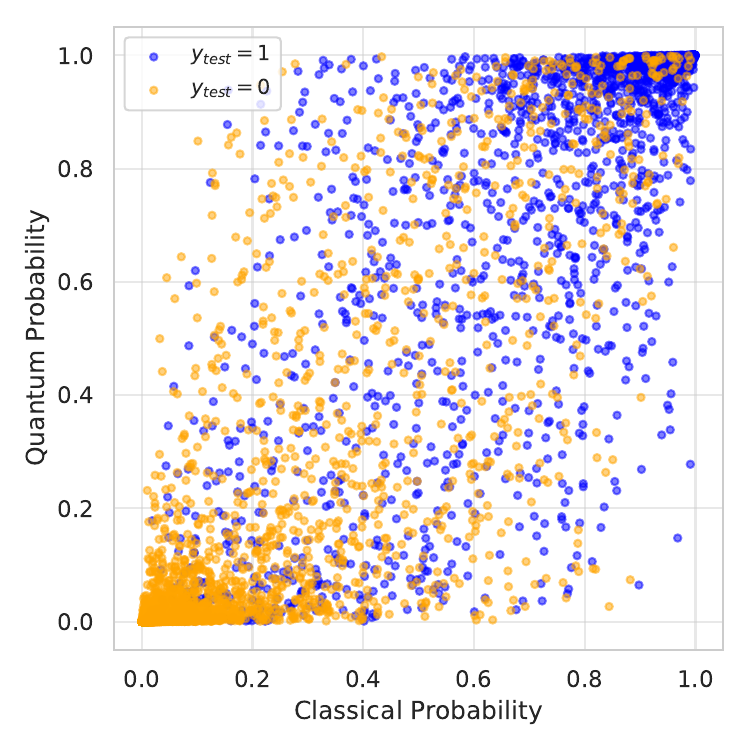} 
  \caption{Diversity between quantum and classical model predictions.
Quantum model probabilities versus classical model probabilities across test dataset points.}
  \label{fig:model_corr}
\end{figure}

In practice, managing this trade-off is key: ensembles must consist of individually strong models that are also complementary. 
Too much homogeneity among base learners limits ensemble benefits, while introducing overly weak learners in the name of diversity can degrade accuracy. 
The most effective ensembles, therefore, balance low bias, low variance, and high diversity.
As shown in Fig.~\ref{fig:model_corr}, the distribution of prediction scores differs substantially between classical and quantum models, which in turn shapes both score diversity.
Although individual quantum and classical models achieve comparable performance, their prediction patterns diverge due to distinct inductive biases and architectural differences. 
This combination of similar accuracy and differing outputs satisfies the core criteria for successful ensemble learning, as formalized in Eq.~\ref{eq:bias_variance_diversity}.
This diversity between a quantum and classical model is quantified by comparing the average feature values for the top $100$ customers (Table ~\ref{tab:feature_comp_unbal}). 
These findings underscore the consumer criteria that are most significant in prediction and offer practical insights for business tactics focused on identifying customers with a high likelihood of credit card default. 
Both classical and quantum models demonstrate sensitivity to default-related factors (designated as `$D\_$’) and the risk-related factors (designated as `$R\_$’), although their relative weighting and impact vary across the two methodologies.

\begin{table}[ht!]
    \centering
    \begin{tabular}{|c|c|c|}
    \hline
    Features & Classical model & Quantum Model  \\
    \hline
D\_103 & 0.560 & 0.650 \\ 
D\_68 & 0.528 & 0.445 \\ 
D\_104 & 0.502 & 0.583 \\ 
R\_15 & 0.370 & 0.440 \\ 
R\_24 & 0.320 & 0.260 \\ 
R\_4 & 0.870 & 0.820 \\ 
D\_54 & 0.226 & 0.276 \\ 
D\_39 & 0.529 & 0.480 \\ 
R\_27 & 0.619 & 0.668 \\ 
D\_58 & 0.489 & 0.535 \\ 
    \hline
    \end{tabular}
    \caption{
    Comparison of the top 20 scaled feature averages, ranked by decreasing average differences, for the top 100 customers selected by classical and quantum models in credit default prediction on an unbalanced dataset.
    }
    \label{tab:feature_comp_unbal}
\end{table}


After validating the results as described in Section~\ref{sec:results}, we proceeded to utilize an unbalanced dataset with $10,000$ samples that maintained a comparable distribution to the entire dataset (about $80\%$ negatives and $20\%$ positives). 
Given this new proportion, it became more suitable to utilize the CDR metric, as its label weights were derived from a distribution that represented the sub-sampling of $5\%$ of the negative class from the entire dataset. 
We investigated the application of Bayesian optimization to determine the hyperparameters of the \texttt{XGBClassifier} in both classical and quantum models to enhance their performance. 
Specifically, we employed 15 distinct random seeds for the \texttt{BayesSearchCV} from Scikit-Optimize~\cite{scikit-optimize} and developed ensemble models for all 225 combinations of the classical and quantum models.


In particular, the \texttt{BayesSearchCV} was performed on the following parameters~\cite{xgboost-docs}: 
\begin{itemize}
    \item \texttt{learning\_rate}: multiplier applied to the step size to prevent overfitting.
    \item \texttt{min\_child\_weight}: minimum sum of instance weights needed in a child node of a tree for a split to be allowed.
    \item \texttt{subsample}: proportion of the data used to train each tree.
    \item \texttt{colsample\_bytree}: proportion of features randomly sampled for constructing each tree.
    \item \texttt{max\_depth}: maximum depth of each tree.
    \item \texttt{gamma}: minimum loss reduction needed to make a further partition on a leaf node of the tree.
    \item \texttt{n\_estimators}: number of trees.
\end{itemize}
The outcomes of these Bayesian models are presented in Table~\ref{tab:final_results}. 

\begin{table}[ht!]
    \centering
    \begin{tabular}{|c|c|c|}
    \hline
     & \multicolumn{2}{|c|}{CDR metric}\\
    \cline{2-3}
    Models & \textbf{MPS Simulator} & \textbf{\textit{ibm\_marrakesh}}  \\
    \hline
    Classical & 0.7972 $\pm$ 0.0029 & 0.7982 $\pm$ 0.0028 \\
    Quantum   & 0.7931 $\pm$ 0.0021 & 0.7822 $\pm$ 0.0043 \\
    Means-model  & \textbf{0.8013 $\pm$ 0.0019} & 0.7991 $\pm$ 0.0027 \\ 
    LogReg ($C=0.2$)  & 0.7987 $\pm$ 0.0034 & 0.7973 $\pm$ 0.0049 \\
    LogReg ($C=0.04$)  & 0.8008 $\pm$ 0.0023 & \textbf{0.8000 $\pm$ 0.0032} \\
    \hline
    \end{tabular}
    \caption{CDR metric from the MPS simulations and the IBM quantum hardware experiments on unbalanced data.}
    \label{tab:final_results}
\end{table}

Two categories of ensemble models were employed in our initial ensembling experiments: mean-based models, wherein the prediction probability is determined as the average of classical and quantum prediction probabilities; and classifier-based models, which utilize the prediction probabilities from classical and quantum models as inputs to train a classical classifier, such as Logistic Regression.
%
%
Although individual quantum models seem to exhibit marginally inferior performance compared to their classical counterparts, the ensemble models surpassed the purely classical models. 
This outcome underscores the potential of integrating quantum-based models with classical approaches to enhance the efficacy of machine learning pipelines.

Additionally, the observed similarity in performance between classical and quantum models suggests that noise present in current quantum hardware does not significantly obscure the informative signal at the scale considered. 
This finding is particularly relevant for near-term quantum devices, where noise and decoherence are often regarded as primary limitations to practical utility. 
The robustness of the quantum feature maps and circuit design against realistic noise levels indicates that these models can retain predictive power without requiring full error correction. 
Such resilience is critical for the development of hybrid quantum-classical workflows, as it supports the feasibility of deploying parameterized quantum circuits for machine learning tasks in the pre-fault-tolerant era. 
Furthermore, these results motivate further exploration of noise-aware training strategies and error mitigation techniques to enhance performance as system sizes and circuit depths increase.

\section{\label{sec:conclusion}Conclusions}

This article examined the efficacy of ensemble models, composed of classical and quantum models, in accurately predicting the probability of credit card default. 
For the restricted set of data (sampled from the original large dataset), the results indicate that both the ensemble means-model and the logistic regression ensemble approaches achieve better performance than the individual classical and quantum models. 
On the unbalanced $10,000$-sample evaluation using the CDR metric, the ensemble improves over the classical baseline by $\Delta=0.0041$ on the MPS simulator (means-ensemble: $0.8013\pm0.0019$ vs.\ classical: $0.7972\pm0.0029$) and by $\Delta=0.0018$ (LogReg, $C=0.04$) and $\Delta=0.0009$ (means-ensemble) on \textit{ibm\_marrakesh}
i.e., gains of approximately $9$ to $41$ basis points.
To refine these findings, further experiments are essential due to the evaluation results and associated errors sharing a similar order of magnitude.

The average performance -- evaluated using the CDR metric, which combines the capture rate at $4\%$ and the Gini coefficient -- consistently yields higher scores with low standard deviation across multiple runs or models, indicating robust and reliable predictive capability.
Logistic regression, evaluated at different regularization strengths ($C = 0.2$ and $C = 0.04$), also delivers competitive performance, matching or surpassing individual classical or quantum model performances. 
These improvements can be attributed to the ensemble effect, where aggregating predictions from a diverse set of models, both quantum and classical, reduces variance and leverages the strengths of each contributing model, thereby mitigating the weaknesses of any single configuration.  
Similarly, the combination of ensemble averaging and the inherent stability of logistic regression leads to higher comparably stable CDR metric across seeds/splits, as reflected in the reported mean and standard deviation values.
The consistency in performance between MPS simulations and quantum hardware executions indicates that the observed gain on the restricted set of data is resilient to device-level noise, suggesting minimal degradation due to hardware-induced errors.

Although the ensemble models have demonstrated promising improvements over classical baselines, their practical utility relative to established classical methods executed on classical hardware (CPUs, GPUs, TPUs) remains unresolved. 
Beyond the potential performance enhancements afforded by sophisticated classical machine learning models, two fundamental research questions remain open.
First, can the projected quantum features derived from pre-fault-tolerant quantum devices be efficiently approximated by classical approaches like Pauli propagation methods as scale increases ~\cite{angrisani2024classically}? 
If so, even superior quantum performance may reduce to quantum-inspired strategies rather than a true quantum advantage.
Second, will these gains persist for expectation values in larger systems, or will noise and error mitigation cost obscure any meaningful signal~\cite{schuster2024polynomial}? 
Answering these questions will determine whether ensemble models, as explored in this study, can provide sustained and practical advantages over classical approaches. 

\section{\label{sec:acknowledgment}Acknowledgment}

The authors would like to thank Vladimir Rastunkov, Adel Dayarian, Brian Quanz, Gines Carrascal, and Jae-Eun Park for insightful discussions and feedback throughout the course of this work.

\section{\label{sec:discaimer}Disclaimer}

This paper is provided solely for informational purposes as an academic contribution by the authors to the research community and does not represent, reflect, or constitute the views, policies, positions, or practices of American Express or its affiliates. Nothing in this paper should be cited or relied upon as evidence of, or support for, the business views, policies, positions, or practices of American Express or its affiliates.

\bibliographystyle{IEEEtran}
\bibliography{Reference}

\end{document}